\documentclass{Interspeech}

\interspeechcameraready

\title{Real-Time Audio-Visual Speech Enhancement Using Pre-trained Visual Representations}

\author[affiliation={1, 2}]{Teng}{Ma}
\author[affiliation={2}]{Sile}{Yin}
\author[affiliation={2}]{Li-Chia}{Yang}
\author[affiliation={2}]{Shuo}{Zhang}

\affiliation{School of Music}{Georgia Institute of Technology}{United States}
\affiliation{Research}{Bose Corporation}{United States}
\email{tengofma@gmail.com, sile\_yin@bose.com, richard\_yang@bose.com, shuo\_zhang@bose.com}

\usepackage{comment}

\usepackage{cite}
\usepackage{multirow}
\usepackage{booktabs}
\usepackage{graphicx}
\begin{document}

\maketitle

\begin{abstract}
    
    Speech enhancement in audio-only settings remains challenging, particularly in the presence of interfering speakers. This paper presents a simple yet effective real-time audio-visual speech enhancement (AVSE) system, RAVEN, which isolates and enhances the on-screen target speaker while suppressing interfering speakers and background noise. We investigate how visual embeddings learned from audio-visual speech recognition (AVSR) and active speaker detection (ASD) contribute to AVSE across different SNR conditions and numbers of interfering speakers. Our results show concatenating embeddings from AVSR and ASD models provides the greatest improvement in low-SNR, multi-speaker environments, while AVSR embeddings alone perform best in noise-only scenarios. In addition, we develop a real-time streaming system that operates on a computer CPU and we provide a video demonstration and code repository. To our knowledge, this is the first open-source implementation of a real-time AVSE system.

\end{abstract}

\section{Introduction}
% \sandra{Multilingual; Generalization / More Robust; Personalized Speech Enhancement; }

% \begin{itemize}
    % \item Speech enhancement - important in video calls \cite{zhu_real-time_2023}, hearing aids, in-car voice pickup solutions etc. (Improved Lite AVSE \cite{chuang_improved_2022, chuang_lite_2020})
    % \item Environments usually more complex than just noise, often interfering speakers present; target speaker might have a low volume, hence low SNR; AOSE usually not enough; AVSE has been popular, esp. with the easy access of video input in a video call, and popularity of cameras in a car.
    % \item AVSE usually studied in a non-causal setting. 
    % \item There are RT-AVSE systems but the real time ones didn't realease code or demo so it's hard to know how well the system works in real time.
    % \item This paper presents XXX
% \end{itemize}

Speech enhancement is essential in various applications, both as a standalone task and as a component of broader speech-related systems. It can improve speech quality in phone calls and hearing devices, as well as enhance downstream tasks such as automatic speech recognition (ASR) and speech-based biometric authentication. In academic research, speech enhancement is commonly defined as the suppression of background noise \cite{das_fundamentals_2021, michelsanti_overview_2021}, which has traditionally been studied in the context of environmental noise \cite{jung_flowavse_2024}. However, real-world acoustic environments are far more complex, where background noise includes not only non-speech sounds but also interference from competing speakers. To address this, Personalized Speech Enhancement has gained increasing attention \cite{eskimez_personalized_2022}, aiming to remove both environmental noise and interfering speech. Traditional audio-only speech enhancement (AOSE) methods often struggle in multi-speaker scenarios unless provided with enrollment audio from the target speaker\cite{zhu_real-time_2023}. This limitation has led to the emergence of audio-visual speech enhancement (AVSE) as a promising alternative, leveraging the growing accessibility of visual inputs on modern devices.

Most existing AVSE work has been conducted in a non-causal setting, employing either mask-based\cite{ephrat_looking_2018, afouras_conversation_2018, michelsanti_overview_2021}, mapping-based\cite{chuang_improved_2022, chuang_lite_2020, ma_end--end_2021}, or synthesis-based approaches\cite{jung_flowavse_2024, yang_audio-visual_2022}. Many systems rely on raw visual input \cite{yang_audio-visual_2022, ephrat_looking_2018, afouras_conversation_2018}, which imposes high computational demands and requires extensive training data. Hence, some work reduces the dimensionality through classical approaches like Discrete Cosine Transform\cite{adeel_novel_2020, adeel_lip-reading_2021} or using facial landmarks\cite{hou_audio-visual_2016}. Alternatively, some work has also explored using extracted embeddings that encode speech-related information in a low-dimensional latent space, such as multisensory features designed for audio-visual synchronization\cite{ferrari_audio-visual_2018}. Similarly, embeddings extracted from related speech tasks have also been proven to be effective\cite{inan_evaluating_2019}. While AVSE methods have shown significantly better performance than AOSE in complex acoustic environments, practical deployment in real-world applications requires a real-time, low-latency operation. Recently, causal AVSE models have been proposed. \cite{gogate_cochleanet_2020} introduced a deep neural network that estimates an ideal binary mask for noise-only settings. \cite{zhu_real-time_2023} proposed a multi-stage gating-and-summation fusion module to merge audio and visual cues. \cite{chen_rt--voce_2024} replaced the Transformer encoder with the Emformer and introduced a causal neural vocoder. However, to our knowledge, these works do not provide publicly available implementations and some work does not disclose the initial signal-to-noise ratio (SNR) conditions of the input mixture, which makes benchmark comparison and reproduction challenging.

To address these gaps, we propose a simple yet effective causal AVSE model that uses pre-trained visual encoders trained on speech-related tasks. Our contributions are summarized as follows: (1) We provide a fully reproducible code base\footnote{See \url{https://github.com/Bose/RAVEN/}} for a real-time audio-visual speech enhancement algorithm, which to our knowledge is the first open-source implementation for a real-time AVSE system that can operate on a CPU. A video demonstration is 
\ifinterspeechfinal
    available in our code repository. 
\else
    in the supplementary material. 
\fi
(2) We conduct an in-depth analysis of how visual embeddings extracted from different audio-visual speech tasks, namely audio-visual speech recognition (AVSR) and audio-visual active speaker detection (ASD), impact AVSE performance.

\section{Methodology}

\subsection{Audio-visual fusion model}
\begin{figure*}[ht]
    \centering
    \includegraphics[width=\textwidth]{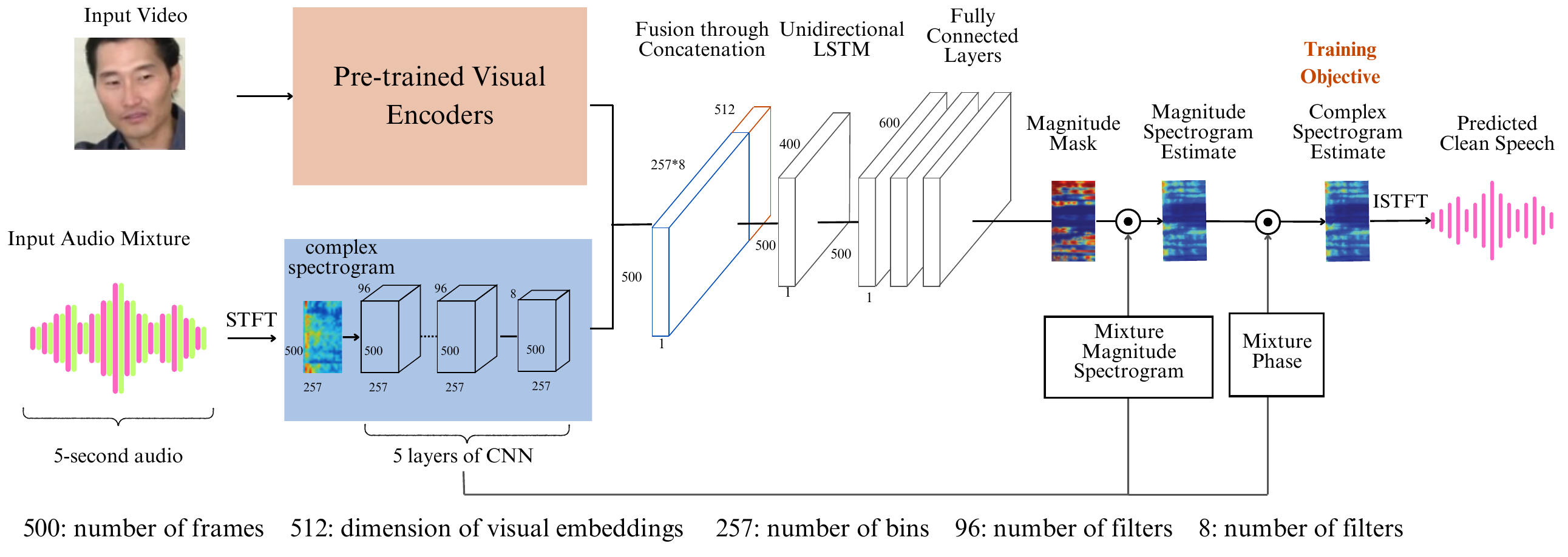}
    \caption{\textbf{Architecture of our mask-based late fusion method.} Each CNN layer in the audio stream is followed by a batch normalization layer and a ReLU layer. The temporal receptive field of the CNN layers is 5, with a 2-frame lookahead. Each fully connected layer is followed by a ReLU layer. A Sigmoid function is applied to the magnitude mask to limit the mask values from 0 to 1.}
    \label{fig:model-arch}
\end{figure*}
As shown in Fig.~\ref{fig:model-arch}, our proposed model is a mask-based late fusion approach \cite{wilson_exploring_2018} that incorporates phase information to improve speech enhancement. The audio stream transforms the raw \SI{16}{\kilo\hertz} input into the time-frequency domain using the Short-Time Fourier Transform (STFT). The resulting spectrograms are passed through a 5-layer convolutional neural network (CNN). The STFT parameters are set as follows: a Hann window of length 400, a hop size of 160, and 512 frequency bins ($n_{fft}$), with a power-law compression rate ($p$) of 0.3, which has been widely adopted in speech enhancement systems \cite{ephrat_looking_2018}. The audio CNN is not causal and has a kernel size of 5 in its first layer and 1 in the rest, resulting in a smaller receptive field that fits in the video lookahead frame mentioned in Section ~2.3.
% \sandra{TODO: If space, could talk about how we chose 5 layers of CNN}

The video stream utilizes the front-end output of a pre-trained visual encoder. All visual embeddings are extracted at a frame rate of 25 frames per second, which is subsequently upsampled to match the frame rate of the audio stream latent embeddings. The outputs of both streams are then concatenated and passed through a unidirectional long short-term memory network, followed by 3 fully connected layers, to estimate the magnitude mask. Finally, the phase information from the input mixture is combined with the magnitude estimate to reconstruct the complex spectrogram, which is then transformed back into the time domain to obtain the enhanced speech signal.

Since the model estimates a magnitude mask, phase information is partially lost. To mitigate this issue, our phase-sensitive loss function (PSA) consists of MSE loss for both the magnitude and the complex spectrogram, inspired by \cite{erdogan_phase-sensitive_2015, wang_compensation_2021}. The overall loss function is formulated as follows,

% \begin{align}
%     \mathcal{L}_{\text{PSA}} &= \text{MSE} \big( \Re(\hat{S}), \Re(S) \big) \notag \\
%     &\quad + \text{MSE} \big( \Im(\hat{S}), \Im(S) \big) \notag \\
%     &\quad + \text{MSE} \big(|\hat{S}|, |S|\big), \label{eq:RI}
% \end{align}

\begin{equation}
    \mathcal{L}_{\text{PSA}} = \Vert \hat{S} - S\Vert^2 + \Vert |\hat{S}| - |S| \Vert^2
\end{equation}
where \(\hat{S}\) and \(S\) are the power-law compressed complex spectrograms of the estimated and target clean speech, with a compression rate $p$ of 0.3, and \(|\cdot|\) extracts the magnitude.

\subsection{Pre-trained visual front end}

We evaluate AVSE using four pre-trained visual front-ends: two for AVSR and two for ASD.

\subsubsection{Audio-visual speech recognition}
For AVSR, we investigate two pre-trained models: AV-HuBERT\footnote{\url{https://github.com/facebookresearch/av\_hubert}} \cite{shi_learning_2022, shi_robust_2022} and Visual Speech Recognition for Multiple Languages in the Wild (VSRiW)\footnote{\url{https://github.com/mpc001/Visual\_Speech\_Recognition\_for\_Multiple\_Languages}} \cite{ma_visual_2022, ma_end--end_2021}. AV-HuBERT extends the HuBERT framework to multimodal inputs, learning joint speech representations from audio and lip movements. The version used in this study was fine-tuned on 433 hours of English video data from LRS3. VSRiW is an end-to-end visual speech recognition model that integrates feature extraction with a hybrid CTC/attention back-end. The version used was trained on GRID, achieving a word error rate (WER) of 4.8. We also experimented a model version that was trained on LRS3, but with a WER of 32.3. 

Both models share the same visual front-end architecture, consisting of a ResNet encoder with a 3D convolutional layer that has a kernel size of 5, padding of 2 on both sides along the time axis. This front-end architecture results in a receptive field of 5 frames, and a 2-frame lookahead. To ensure comparability, we follow \cite{ma_visual_2022} to extract the same mouth region. Using MediaPipe, we detect 68 facial landmarks and align faces to a neutral reference frame via similarity transformation to remove translation and scaling variations. It is then cropped by a 96×96 bounding box centered on the mouth, converted to grayscale, and normalized based on the training set's mean and variance. 

\subsubsection{Active speaker detection}
For ASD, we investigate TalkNet\footnote{\url{https://github.com/TaoRuijie/TalkNet-ASD}} \cite{tao_is_2021} and LoCoNet\footnote{\url{https://github.com/SJTUwxz/LoCoNet_ASD}} \cite{wang_loconet_2024} because these models may encode mapped information between facial identity and speech activity. TalkNet consists of audio and visual temporal encoders, an audio-visual cross-attention mechanism for intermodality interaction, and a self-attention mechanism for long-term speaking evidence. LoCoNet builds on TalkNet by incorporating long-term intra-speaker and short-term inter-speaker modeling.

Similar to speech recognition models, both ASD models employ a ResNet encoder with a 3D convolutional layer with identical kernel size, padding, receptive field, and lookahead. However, after being converted to grayscale, the ASD models crop a 112×112 face region before normalization.  

All embeddings have a dimension of 512, except for AV-HuBERT, which has a dimension of 768.

\subsubsection{Combined tasks}

Combining pre-trained feature extractors has proven to be effective in language\cite{peters_deep_2018}, vision\cite{chowdhury_few-shot_2021}, and audio modalities\cite{ma_music_2024}. In natural language processing specifically, concatenating multiple pre-trained contextualized embeddings, traditional non-contextualized embeddings, and character embeddings has been proved to further improve various prediction tasks\cite{peters_deep_2018, akbik_contextual_2018}. Inspired by this, we have also experimented with concatenating the best performing visual front-end for each task.

\subsection{Real-time system implementation}
We implemented a real-time system that streams audio-visual input at 25 frames per second on an Apple M3 Max SOC. The pre-trained visual encoders used in our model have a receptive field of 5 video frames with a 2-frame lookahead, requiring a buffer of 5 frames to produce a single meaningful visual embedding of the middle (3rd) frame. This introduces an algorithmic latency of \SI{120}{\milli\second} (\SI{40}{\milli\second} from current frame plus  \SI{80}{\milli\second} from 2-frame lookahead), despite our fusion model operating on a frame-by-frame basis. Our system meets the real-time requirement for all pre-trained embeddings listed above, since the processing time per frame remains below the streaming frame interval of \SI{40}{\milli\second}. Our demo video uses the VSRiW embeddings, which yields a processing time of \SI{28}{\milli\second} per frame.
% \sandra{could make graphic for this too; add language about the actual system using AVSR and report processing latency on that}

\section{Experiment setup}

\subsection{Datasets} \label{sec:datasets}

We use VoxCeleb2 as the speech dataset and MUSAN for noise and music. VoxCeleb2 is an audio-visual speech dataset containing approximately one million utterances from various celebrities, extracted from videos uploaded to YouTube recorded in diverse acoustic environments. The dataset is multilingual and includes utterances of variable lengths. For both training and testing, we use 5-second video clips, padding shorter clips as necessary. The audio sample rate is \SI{16}{\kilo\hertz}, while the video has a frame rate of 25 frames per second. The train/validation/test split for both VoxCeleb2 and MUSAN is included in our code repository.

To simulate a realistic speech enhancement scenario where background interference often consists of more than just noise, we generate input mixtures by randomly selecting one interfering speaker from VoxCeleb2. Additionally, we introduce a second interfering source, which is randomly selected as another speaker, music, noise, or no additional interference. To ensure our model performs in a challenging evaluation setting, we constrain the signal-to-noise ratio (SNR) of the input mixtures to a range of \SI{-10}{\decibel} to \SI{10}{\decibel}.  Since VoxCeleb2 is an in-the-wild dataset, we generate ground truth by denoising the original audio using DeepFilterNet\cite{schroter_deepfilternet_2022}, which produces reasonably clean speech for most utterances upon inspection.

To improve model robustness, we apply video augmentation using three masking techniques: average masking, duplicate frame masking, and zero frame masking. Each method masks up to 10 consecutive frames. Average masking replaces them with the mean visual embedding of the clip, duplicate frame masking uses the preceding frame, and zero frame masking uses the output of the visual encoder when the input consists of all zeros. This strategy enhances generalization to scenarios with occluded or missing visual information.

\subsection{Visual encoder comparison study}

We begin by comparing the performance of visual encoders trained on the same task to identify the stronger-performing model. Next, we assess whether combining the two pre-trained visual embeddings from the same task enhances speech enhancement performance. We then investigate whether integrating the best-performing visual embedding from each task, namely AVSR and ASD, yields further improvements.  

To evaluate speech quality, we use the wide-band Perceptual Evaluation of Speech Quality (PESQ-WB)\cite{rix_perceptual_2001}, Scale-Invariant Signal-to-Distortion Ratio (SISDR)\cite{roux_sdr_2019}, and Extended Short-Time Objective Intelligibility (ESTOI)\cite{jensen_algorithm_2016}. We compare these test metrics in three test conditions: noise only, one interfering speaker, and three interfering speakers.

\subsection{Generalization across datasets}
To evaluate the generalizability of our model across different languages, scenarios, acoustic environments, and facial expressions, we test our best-performing models on two additional audio-visual datasets alongside VoxCeleb2: ViSPeR\cite{narayan_visper_2024} and MEAD\cite{vedaldi_mead_2020}. ViSPeR contains YouTube-sourced videos of speech in Chinese, Spanish, Arabic, and French with minimal background noise. MEAD, in contrast, is a lab-controlled dataset featuring 60 actors and actresses speaking with eight different emotions at three intensity levels. For evaluation, we randomly select 1000 samples and apply the same audio mixture processing pipeline from Section \ref{sec:datasets}, combining target speech with interfering speakers from VoxCeleb2 and background noise from MUSAN. 

\section{Results and discussion}
\subsection{Visual encoder performance analysis}

\begin{table}[t]
    \centering
    % \scriptsize 
    \caption{Model performance in noise-only scenario with mixed SNR condition from [\SI{-10}{\decibel}, \SI{10}{\decibel}]. Model with VSRiW as visual embedding performs the best in all noise-only settings.}
    \label{tab:noise_only_mixed_SNR}
    \begin{tabular}{l lcccl}
        \toprule
        \textbf{Task} & \textbf{Visual Encoder} & \textbf{PESQ} & \textbf{SISDR} &  \textbf{ESTOI}\\
        \midrule
        \multirow{3}{*}{AVSR} & \textbf{AVH}uBERT & 2.057 & 8.980 & 0.775\\
                              & \textbf{V}SRiW    & \textbf{2.063} & \textbf{9.239} & \textbf{0.779}\\
                              & \textbf{AVH}+\textbf{V} & 2.044 & 8.967 & 0.776 \\
        \midrule
        \multirow{2}{*}{ASD}  & \textbf{T}alkNet  & 2.048 & 9.062 & 0.777\\
                              & LoCoNet  & 1.923 & 7.318 & 0.742\\
        \midrule
        \multirow{1}{*}{Both} & \textbf{AVH}+\textbf{T} & 2.032 & 8.924 & 0.775\\
        \bottomrule
    \end{tabular}
\end{table}

\begin{table*}[t]
    \centering
    % \scriptsize 
    % \renewcommand{\arraystretch}{1.05} 
    \caption{Model performance under different SNR levels with One and Three Interfering Speakers. Model with AVSR visual embedding and ASD visual embedding concatenated (AVHuBERT+TalkNet) performs the best in low-SNR, multi-speaker settings.}
    \label{tab:multi_speaker_perform}
    \resizebox{\textwidth}{!}{ 
    \begin{tabular}{l|ccc|ccc|ccc|ccc}
        \toprule
        \multirow{4}{1.5cm}{\centering \textbf{Visual Encoder}} 
        & \multicolumn{6}{c|}{\textbf{One Interfering Speaker}} 
        & \multicolumn{6}{c}{\textbf{Three Interfering Speakers}} \\ 
        \cmidrule(lr){2-7} \cmidrule(lr){8-13}
        & \multicolumn{3}{c|}{\textbf{\SI{-10}{\decibel}}} & \multicolumn{3}{c|}{\textbf{\SI{-5}{\decibel}}} 
        & \multicolumn{3}{c|}{\textbf{\SI{-10}{\decibel}}} & \multicolumn{3}{c}{\textbf{\SI{-5}{\decibel}}} \\ 
        \cmidrule(lr){2-4} \cmidrule(lr){5-7} \cmidrule(lr){8-10} \cmidrule(lr){11-13}
        & PESQ & SISDR & ESTOI & PESQ & SISDR & ESTOI 
        & PESQ & SISDR & ESTOI & PESQ & SISDR & ESTOI \\ 
        \midrule
        \textbf{AVH}uBERT  & 1.228 & 0.449 & 0.543 & 1.401 & 3.766 & 0.646 
                  & 1.097 & -5.018 & 0.328 & 1.179 & -0.286 & 0.458 \\
        \textbf{V}SRiW     & 1.215 & -0.006 & 0.534 & 1.378 & 3.591 & 0.639 
                  & 1.093 & -5.568 & 0.316 & 1.176 & -0.335 & 0.454 \\
        \textbf{AVH+V}     & 1.231 & 0.536 & 0.549 & 1.403 & 3.797 & 0.648 
                  & 1.099 & -4.908 & 0.332 & 1.184 & -0.255 & 0.460 \\
        \midrule
        \textbf{T}alkNet   & 1.226 & 0.307 & 0.542 & 1.387 & 3.655 & 0.642 
                  & 1.100 & -5.109 & 0.323 & 1.175 & -0.435 & 0.449 \\
        LoCoNet   & 1.146 & -2.884 & 0.453 & 1.261 & 1.339 & 0.568    
                  & 1.074 & -8.075 & 0.237 & 1.117 & -2.749 & 0.364 \\
        \midrule
        \textbf{AVH+T}     & \textbf{1.231} & \textbf{0.600} & \textbf{0.550} & 1.401 & \textbf{3.882} & \textbf{0.649} 
                  & 1.096 & \textbf{-4.762} & \textbf{0.336} & 1.179 & \textbf{-0.174} & \textbf{0.461} \\
        \bottomrule
    \end{tabular}
    }
\end{table*}

\begin{table*}[t]
    \centering
    % \scriptsize 
    % \renewcommand{\arraystretch}{1.05} % Adjust row spacing for readability
    \caption{\textit{Model performance with VSRiW as visual embedding across datasets in SNR Scenarios randomly sampled from [\SI{-10}{\decibel}, \SI{10}{\decibel}]. Our approach generalizes to diverse languages but experiences performance degradation on various speaker emotions.}}
    \label{tab:dataset_general}
    % \resizebox{\textwidth}{!}{ % Auto-scale table to fit the page width
    \begin{tabular}{l|ccc|ccc|ccc}
        \toprule
        \multirow{2}{*}{\textbf{Dataset}} 
        & \multicolumn{3}{c|}{\textbf{Noise Only}} 
        & \multicolumn{3}{c|}{\textbf{One Interfering Speaker}} 
        & \multicolumn{3}{c}{\textbf{Three Interfering Speakers}} \\ 
        \cmidrule(lr){2-4} \cmidrule(lr){5-7} \cmidrule(lr){8-10}
        & PESQ & SISDR & ESTOI & PESQ & SISDR & ESTOI 
        & PESQ & SISDR & ESTOI \\ 
        \midrule
        VoxCeleb2  & 2.063 & 9.239 & 0.779 & 1.688 & 6.184 & 0.713 & 1.499 & 3.944 & 0.597 \\
        \midrule
        ViSPeR     & 1.912 & 8.824 & 0.773 & 1.603 & 5.739 & 0.712 & 1.414 & 3.448 & 0.589 \\
        MEAD       & 1.545 & 3.528 & 0.612 & 1.285 & 0.729 & 0.510 & 1.226 & -1.803 & 0.396 \\
        \bottomrule
    \end{tabular}
    % }
\end{table*}

We present our results as follows: Table~\ref{tab:noise_only_mixed_SNR} presents the performance of our models with different visual encoders in a noise-only scenario with SNR uniformly sampled from [\SI{-10}{\decibel}, \SI{10}{\decibel}]. Table~\ref{tab:multi_speaker_perform} reports performance in low-SNR, multi-speaker settings. All fusion models were trained for 5 epochs and have 6 million parameters.
\subsubsection{Same task comparison}
For visual embeddings trained on AVSR, we observe that AV-HuBERT performs better in low-SNR conditions with interfering speakers, while VSRiW is more effective in noise-only scenarios. This may be due to differences in how noise was incorporated during training. AV-HuBERT was trained with a diverse set of noise types, including natural sounds, music, babble, and speech at \SI{0}{\decibel}, which may have better prepared it for multi-speaker environments. In contrast, VSRiW's noise augmentation involved only babble noise, mixed at SNR levels uniformly sampled from [\SI{-5}{\decibel}, \SI{20}{\decibel}], which could explain its stronger performance in noise-only conditions but weaker performance in multi-speaker scenarios. Concatenating embeddings from models trained on the same task yields only marginal improvements, with the most noticeable benefits appearing in low-SNR conditions with multiple interfering speakers, as shown in Table~\ref{tab:multi_speaker_perform}. This suggests that additional speaker-related information from the same task can help disentangle overlapping speech signals. Additionally, the version of VSRiW trained on LRS3 underperforms across all AVSE test conditions, aligning with its weaker WER in AVSR.

For visual embeddings trained on ASD, TalkNet consistently outperforms LoCoNet in all test conditions. One possible explanation is that LoCoNet incorporates short-term inter-speaker modeling, whereas TalkNet focuses on long-term intra-speaker modeling. Since AVSE primarily depends on tracking the target speaker’s speech activity rather than modeling interactions between multiple speakers, TalkNet’s embeddings likely provide more relevant information for the enhancement task. Given that both models produce embeddings of the same dimension, LoCoNet’s weaker performance suggests that its features may be less useful for speech enhancement.

\subsubsection{Cross task comparison}
Comparing AV-HuBERT and TalkNet in the AVSE task, we observe that AV-HuBERT performs better in low-SNR, multi-speaker environments. This suggests that continuous lip movement information from AVSR is more beneficial for AVSE in such conditions than the segmented activity information provided by ASD. In addition, combining visual embeddings from AVSR and ASD models improves the fidelity and intelligibility of clean speech, particularly in low-SNR, multi-speaker environments. This is reflected in gains in SISDR and ESTOI metrics, indicating that incorporating target speaker activity cues enhances speech separation.

Overall, AVSR visual encoders are more effective for speech enhancement in noise-only scenarios, while in low-SNR, multi-speaker environments, concatenating AVSR and ASD embeddings leads to better fidelity and intelligibility.

\subsection{Generalization across datasets}

We report the performance of our model on various datasets in Table~\ref{tab:dataset_general}. Since the goal of this section is to evaluate how well our approach generalizes to real-world scenarios using different datasets, we maintain the same pre-trained visual encoder used in our real-time system implementation: VSRiW trained on the GRID dataset. Our results show that the model generalizes well to in-the-wild datasets, particularly across different languages, as demonstrated by its performance on ViSPeR. However, when tested on MEAD, which includes intense and less common emotional expressions, performance declines. This is likely because extreme emotions introduce variations in lip movements, speech quality, and facial expressions that are rarely seen in publicly available YouTube videos or celebrity interviews, which is what most AVSE work is trained on. This calls on future work to improve robustness to variations in speaker emotions and facial expressions.

\section{Conclusion}
In this work, we propose a simple yet effective phase-aware audiovisual fusion network that leverages visual embeddings from other speech tasks for speech enhancement. We systematically analyze how visual embeddings from audio-visual speech recognition (AVSR) and active speaker detection (ASD) impact AVSE performance. Our results indicate that concatenating AVSR and ASD embeddings improves speech enhancement in low-SNR, multi-speaker conditions, while AVSR embeddings alone perform best in noise-only environments. Additionally, we evaluate our model's generalization across datasets covering different languages, acoustic environments, and facial expressions. We also describe the implementation of our real-time system and provide a demonstration of its practical use.

% \ifinterspeechfinal
%      The Interspeech 2025 organisers
% \else
%      The authors
% \fi
% would like to thank ISCA and the organising committees of past Interspeech conferences for their help and for kindly providing the previous version of this template.

\bibliographystyle{IEEEtran}
\bibliography{references}

\end{document}